\documentclass[12pt,preprint]{aastex}
\renewcommand{\cite}{\citealp}

\newcommand{\rrl}{{RR~Lyrae}}

\shorttitle{Variable stars in the Coma dSph}
\shortauthors{Musella et al.}

\begin{document}


\title{Pulsating Variable Stars in the Coma Berenices dwarf spheroidal galaxy\altaffilmark{1}}


\author{Ilaria Musella\altaffilmark{2},
Vincenzo Ripepi\altaffilmark{2},
Gisella Clementini\altaffilmark{3},
Massimo Dall'Ora\altaffilmark{2},
Karen Kinemuchi\altaffilmark{4},
Luca Di Fabrizio\altaffilmark{5},
Claudia Greco\altaffilmark{6},  
Marcella Marconi\altaffilmark{2},
Horace A. Smith\altaffilmark{7},
Mario Radovich\altaffilmark{2}
and 
Timothy C. Beers\altaffilmark{8}
}

\altaffiltext{1}{Based on data collected at the 1.52 m telescope of the INAF-Osservatorio Astronomico di Bologna, Loiano, 
Italy, at the 2.3 m telescope at the Wyoming Infrared Observatory (WIRO) at Mt. Jelm, 
Wyoming, USA, and at 2.5 m Isaac Newton Telescope, La Palma, Canary Island, Spain.}
\altaffiltext{2}{INAF, Osservatorio Astronomico di Capodimonte, Napoli, Italy,
ilaria@na.astro.it, ripepi@na.astro.it, dallora@na.astro.it, marcella@na.astro.it,radovich@na.astro.it}
\altaffiltext{3}{INAF, Osservatorio Astronomico di
Bologna, Bologna, Italy; gisella.clementini@oabo.inaf.it}
\altaffiltext{4}{Universidad de Concepci\'on, Departamento de
F\'{\i}sica, Concepci\'on, Chile, and University of Florida, Department of
Astronomy, Gainesville, FL 32611-2055, USA;
kkinemuchi@astro-udec.cl}
\altaffiltext{5}{INAF, Centro Galileo Galilei \& Telescopio Nazionale Galileo, S.
Cruz de La Palma, Spain; difabrizio@tng.iac.es}
\altaffiltext{6}{Observatoire de Geneve, Sauverny, Switzerland; claudia.greco@unige.ch}
\altaffiltext{7}{Department of Physics and Astronomy, Michigan State University, East Lansing, 
MI 48824-2320, USA; smith@pa.msu.edu}
\altaffiltext{8}{Department of Physics and Astronomy, CSCE, Center for the Study of Cosmic Evolution, and 
JINA, Joint Institute for Nuclear Astrophysics, Michigan State University, East Lansing, MI 48824, USA;
beers@pa.msu.edu}


\begin{abstract} We present 
$B, V, I$ time-series photometry of the Coma
Berenices dwarf spheroidal galaxy, a faint Milky Way satellite, recently discovered by the Sloan Digital Sky Survey. 
We have obtained $V, B-V$ and
$V, V-I$ color-magnitude diagrams that reach $V\sim 23.0-23.2$ mag showing
the galaxy turnoff at $V\sim 21.7$ mag, and have 
performed the first study of the variable star population of this new Milky Way companion. 
Two RR Lyrae stars (a
fundamental-mode -RRab- and a first overtone -RRc- pulsator) and a short period 
variable with period $P$=0.12468 days were identified in the galaxy.  
The RRab star has a rather long period of $P_{ab}=0.66971$ days and is about 0.2 mag
brighter than the RRc variable and other non-variable stars on the galaxy horizontal branch.
In the period-amplitude diagram the RRab variable  
falls closer to the loci of Oosterhoff type-II systems and 
evolved fundamental-mode RR Lyrae stars in the Galactic globular cluster M3.  
The average apparent magnitude of 
the galaxy horizontal branch, $\left<V_{\rm HB}\right> =18.64 \pm
0.04$ mag, leads to a distance modulus for the Coma dSph $\mu_{0}=18.13 \pm 0.08$ mag, corresponding
to a distance $d$=42$^{+2}_{-1}~\rm{kpc}$, by adopting a reddening $E(B-V) = 0.045 \pm 0.015$ mag
and a metallicity [Fe/H]=$-2.53 \pm 0.05$ dex. 
\end{abstract}





\section{Introduction}
Over the past few years the analysis of the Sloan Digital Sky Survey (SDSS) data 
led to the discovery of  
several ultra-faint companions of the Milky Way (MW). The new systems include two very faint globular clusters (GCs):
Koposov 1 and 2 \citep{kop07}, and fourteen dwarf spheroidal (dSph) galaxies:
Willman\,I, Ursa Major\,I,  Canes Venatici\,I (CVn\,I), Ursa Major\,II (UMa\,II),  
Bootes\,I, Coma Berenices (Coma), Segue\,I, Canes Venatici\,II (CVn\,II), Leo\,IV,  Hercules, Leo\,T, Bootes\,II, Leo\,V, 
and Bootes\,III  
(\citealt{will05a,will05b, zu06a,gr06,zu06b,be06,be07,irw07,wal07,be08,gr08}). The new dSphs have half-light radii similar to those of 
the classical MW dSph companions, however, they are much fainter, with typical effective surface
brightnesses in the range from
28 to 31 arcsec$^{-2}$ (\citealt{be07}, hereafter B07).  
The new discoveries along with the 10 previously known MW dSph companions 
(Draco, Ursa
Minor, Fornax, Carina, Sculptor, Leo\,I, Leo\,II, Sextans, Sagittarius and Canis Major;
\citealt{mateo98}; Ibata, Gilmore, \& Irwin 1995; \citealt{martin04}), bring to twenty-four the number of dSph galaxies presently known to surround the
MW. In the absolute magnitude versus half-light radius plane 
the new dSphs are well separated from both GCs and traditional dSph galaxies, since they are generally 
more extended than GCs and much fainter than the old dSphs 
(see Fig. 8 of B07). Only CVnI, the brightest of the new SDSS dSphs, lies 
on the faintest tail of the distribution for traditional dSphs.   
All the new SDSS dSphs host a metal-poor and ancient  
stellar component, with metallicity generally below [Fe/H]$\sim -2$ dex and down to as low as [Fe/H]$\sim -3$ dex   
(\citealt{mu06,sg07,ge08,ky08}), and age as old as the stars in Galactic GCs like M92 
(see B07).
The density contours are irregular since most of these galaxies are in process of disruption. The 
closest ones (Bootes\,I, UMa\,II and Coma) are more distorted owing to the
 tidal interaction with the MW \citep{be06,be07,zu06b};
others appear to be connected with tidal streams. 
With the exception only of Leo\,T, the new systems all have globular-cluster-like CMDs, showing  main sequences, turn-offs, as well as hints of red 
giant and horizontal branches. 
Variable stars have been identified so far in three of the new systems, namely
Bootes\,I, CVn\,I, and CVn\,II. 
Bootes\,I \citep{dall06,s06} and CVn\,II \citep{greco08} both were found to contain RR Lyrae stars with properties resembling those of 
Oosterhoff-type II \citep{oo39} GCs in the MW. Confirming its similarity to the 
traditional dSphs, CVn\,I \citep{kue08} hosts, instead, RR Lyrae stars with properties intermediate between
the two Oosterhoff types, along with a few candidate Anomalous Cepheids. 

The Coma galaxy ($R.A. = 12^{\rm h} 26^{\rm m} 59.0^{\rm s}$, $DEC=+23^{\circ} 54^{\prime} 15^{''}$, J2000.0, $l$=241.9$^{\circ}$, 
$b$=83.6$^{\circ}$, B07),
located at a heliocentric
distance of $44\pm4$~kpc with a position angle of $120^{\circ}$ and an
absolute magnitude of $M_V = -3.6\pm0.6$~mag, is among the faintest of the SDSS new discoveries. It has half light radii:  
$ r_h (Plummer)$ = 5.0 arcmin, and $ r_h (exponential)$ = 5.9 arcmin, respectively, 
and an irregular and extended shape (B07). 
The $i$, $g-i$ CMD obtained 
 by B07 goes down to $i \sim 25$ mag,  
is consistent with a single, old
stellar population of metallicity $[Fe/H] \sim -2$ , and
shows a poorly populated horizontal branch (HB). 
Metallicities of radial velocity members of the Coma dSph were obtained by \citet{sg07} and 
confirm the low metal abundance 
of the galaxy. These metallicities were re-evaluated by \citet{ky08}, who derived for the galaxy an average 
metallicity $\langle {\rm [Fe/H]}\rangle = -2.53 \pm 0.05$ dex with dispersion 
$\sigma_{\rm [Fe/H]}$=0.45 dex.

In this {\em Letter} we present the first study of the variable stars in the Coma dSph, which led us to  
identify two RR Lyrae stars and one short period 
 variable in the galaxy. We provide multiband light curves 
and properties of 
 the variables,  and use the magnitude of the HB stars to measure the distance to Coma.  We
also present $B$, $B-V$ and $V$, $V-I$ CMDs 
to $V\sim23.0-23.2$ mag, which show the galaxy main
sequence turnoff at $V\sim21.7$ mag, and reveal the presence of several
candidate blue straggler stars (BSSs). 

\section[]{Observations and Data Reduction} 

Time-series $B,V,I$ photometry of the Coma dSph galaxy was collected
on 2007, March 12-16 and April 13-17, with BFOSC at the 1.52 m
telescope of the Bologna Observatory in
Loiano\footnote{http://www.bo.astro.it/loiano/index.htm} ($V$ and $I$
data), on April 30 and May 9-10, using WIRO-Prime, the prime focus
CCD camera (\citealt{p02}) of the 2.3~m Wyoming Infrared Observatory
telescope (WIRO) ($B$ and $V$ data), and $B,V,I$ observations were
gathered on April 21-22, using WFC (Wide Field Camera), the prime
focus mosaic CCD camera of the 2.5 m Isaac Newton Telescope (INT). The
field of view (FOV) covered by the three instruments is $13 \times 12.6$ arcmin$^2$ for BFOSC@LOIANO, $17.8
\times 17.8$ arcmin$^2$ for WIRO, and $33 \times 33$ arcmin$^2$ for
WFC@INT.  We needed two partially overlapping fields 
to cover
the galaxy with BFOSC, while just one WIRO pointing was sufficient, and from the WFC@INT field we could also infer some additional information about the
contamination by field stars and background galaxies. We obtained 14 $B$ images,  
83 $V$  frames 
and 64 $I$ frames 
of the Coma dSph.
Images were pre-reduced following standard procedures (bias
subtraction and flat-field correction) with IRAF.
The $I$-band images were corrected for fringing by adopting,
for each instrument, a well-suited fringing map. We then performed
PSF photometry using the DAOPHOT\,IV/ALLSTAR/ALLFRAME packages
(\citealt{st87,st94}). Typical internal errors of the single-frame photometry for stars at
the HB magnitude level are of about 0.005 mag in all the three bands.
The absolute photometric calibration was derived using observations of standard stars in the Landolt fields SA 101, SA
107, SA 110 and PG1323 (\citealt{la92}), as extended by
P.B. Stetson\footnote{see http://cadcwwwdao.nrc.ca/standards.},
which were obtained at the INT during the night of 2007, April
22. Errors of the absolute photometric calibration are
$\sigma_B=0.01$, $\sigma_V=0.01$ and $\sigma_I=0.02$ mag, respectively.

\section[]{Identification of the Variable Stars}

\begin{figure} 
\includegraphics[width=16.3cm,bb=43 170 570 460,clip]{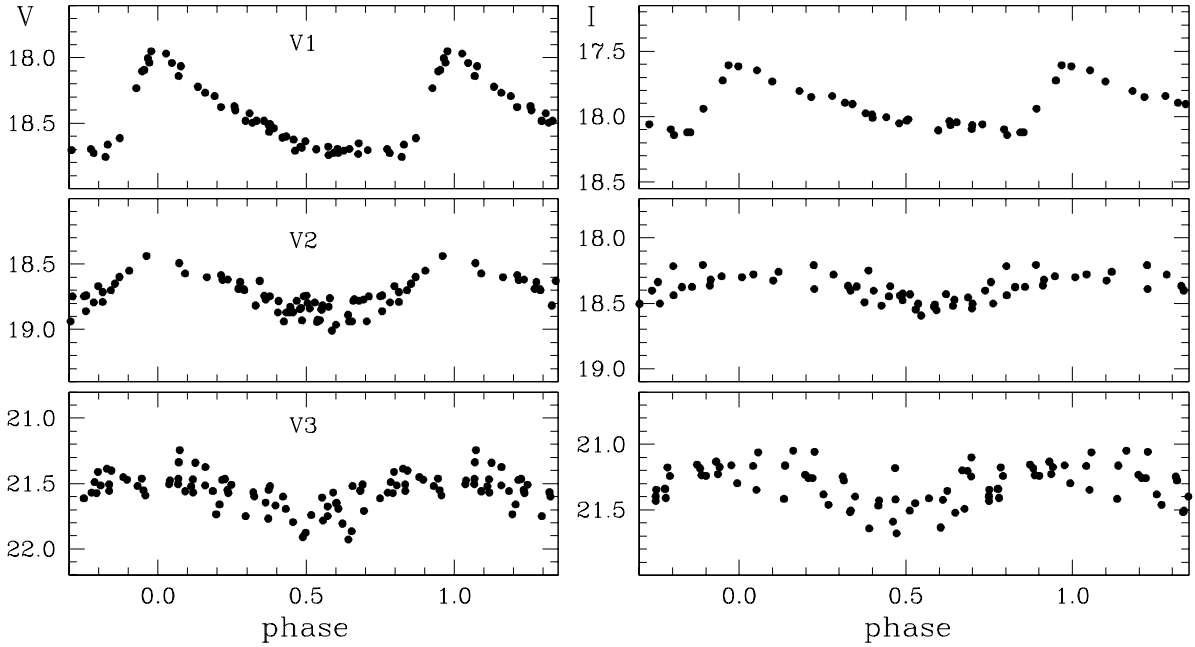}
\caption{$V$ (left panels) and $I$
(right panels) light curves of variable stars discovered in the Coma dSph. From top to 
bottom: fundamental-mode RR Lyrae star, 
first-overtone RR Lyrae pulsator, short period variable.}
\label{f:fig1}
\end{figure}

Variable stars were identified using the $V$ and $I$ time-series data, separately. First we calculated the Fourier transforms (in the
\citealt{sc96} formulation) of the stars having at least 12 measurements in 
each photometric band, then we averaged these transforms to estimate
the noise and calculated the signal-to-noise ratios. Results from the 
$V$ and $I$ photometries were cross-correlated, and all stars with 
S/N$>$ 4 in both photometric bands were visually inspected, for a total of about 
 500 candidates. 
We also checked whether some of the stars in the
BSSs region might be pulsating variables of SX Phoenicis (SX Phe) type.  Study of the
light curves and period derivation were carried out using GRaTiS
(Graphical Analyzer of Time Series, \citealt{clementini00}).  We confirmed the variability and
obtained reliable periods and light curves for 2 RR Lyrae stars: 1 fundamental-mode (RRab)
variable with period $P=0.66971$ days (V1), 1 first-overtone (RRc) pulsator 
with period $P=0.31964$ days (V2), and for one short period variable with period $P=0.12468$ days (V3).
Identification and
properties of the confirmed variable stars are summarized in
Table~\ref{t:table1}; their light curves are shown in
Fig~\ref{f:fig1}. Time-series data for the variable stars are available on request from the first author.  
Both the short period variable and the RRc star lie
inside the half-light radius of the Coma galaxy. 
The RRab lies, instead, slightly outside this region. 
 In the CMD, the short period variable is located in the region of the BSSs, thus suggesting it might be an SX Phe star.
However, 
the star does not fit the SX Phe period-luminosity relations (e.g. \citealt{santolamazza01},
\citealt{poretti08}) unless a pulsation period of one third the value supported by the period search analysis is
adoped. The shape of the light curve and the similarity of the
$A_V,A_I$ amplitudes may suggest
a classification as a W UMa binary system. However, the present data do not allow us 
to definitely assess the actual nature of the star.

The two RR Lyrae stars both fall near the HB of the Coma dSph. However, 
while the average magnitude of the RRc star is fully consistent with the average luminosity of the
HB we infer, by fitting the galaxy CMD with the ridge lines of the Galactic GC M\,68 (see Section 4),  
that the RRab star is about 0.2 mag brighter. The overluminosity of the 
RRab star
could be due to geometrical projection effects, since the
galaxy appears to be elongated due to the tidal interaction with the 
MW (see \citealt{dall08} and 
Section 4). 
Contamination by an
unresolved companion
may also be possible, but this is constrained by the large amplitude of the RRab
star, and by its color,
which is consistent with the expected color of an RRab star. 
A further possibility is that evolution
off the zero-age HB contributes to the enhancement of the brightness of the RRab star (see end of Section 4).

\begin{figure} 
\includegraphics[width=16.3cm]{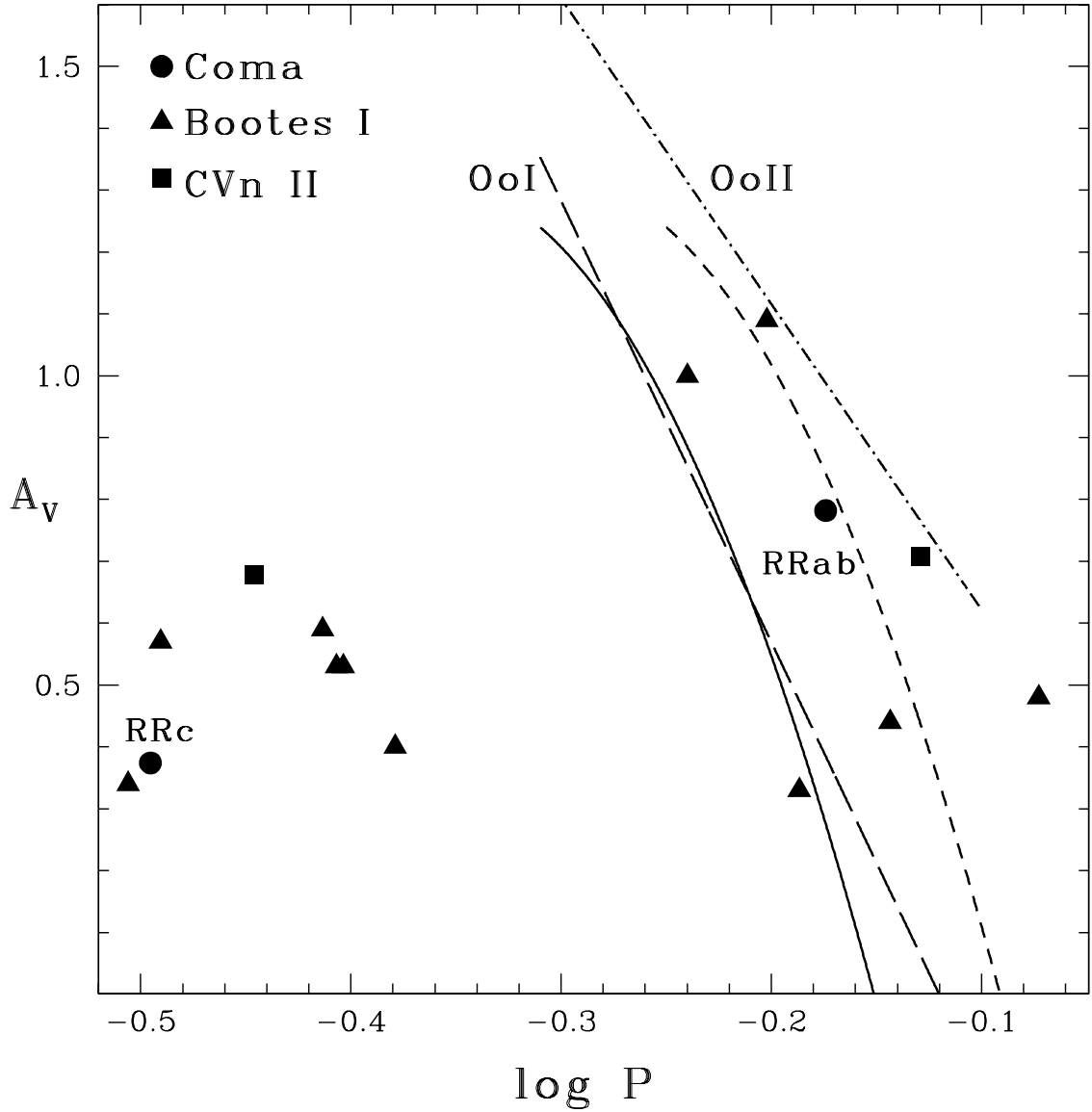}
\caption{Period-amplitude diagram in the $V$ band.
Dashed and dotted-dashed lines  
are the positions of the Oo\,I and Oo\,II Galactic GCs, from \citet{cr00}. 
Period-amplitude distributions of the
{\it bona fide} regular ({\em solid curve}) and well-evolved ({\em dashed curve}) 
fundamental-mode \rrl\ stars in M3, from \citet{cacciari05}, are also shown for comparison.}
\label{f:fig2}
\end{figure}

 The rather long period of the RRab star
suggests that Coma is, like  CVn\,II \citep{greco08}, more similar to 
Oosterhoff type-II systems in the MW, and  
differs from CVn\,I, the brightest of the SDSS dSph's, which has
Oosterhoff-intermediate type (\citealt{kue08}). 
 Fig~\ref{f:fig2} shows the position of the Coma RR Lyrae stars in the $V$-band 
period-amplitude diagram of the Bootes\,I and CVn\,II RR Lyrae stars.  
The Coma RRab star (filled circle) falls closer to the loci of Oo\,II systems (from 
\citealt{cr00}, dashed-dotted line) and evolved fundamental-mode RR Lyrae stars in the Galactic GC M3
(from \citealt{cacciari05}). The RRc star
falls at the short extreme of the bell-shaped distribution defined by Oo\,II RRc stars.
Both these results suggest an Oo\,II classification for the Coma dSph, although with
some caution given the very small number of variable stars.   
We have used the parameters of the Fourier decomposition of the $V$-band light curves, along with
Jurcsik \& Kov\'acs (1996) 
and 
\citet{mor07} formulas, 
to estimate
individual metallicities for the Coma RR Lyrae stars, on the \citet{zw84} metallicity scale. 
Results are summarized in the last column of Table~1, they confirm the low metallicity 
derived for the Coma dSph by the spectroscopic studies.

\section[]{CMD, Structure and distance}
\begin{figure} 
\includegraphics[width=16.8cm,bb=2 100 593 680,clip]{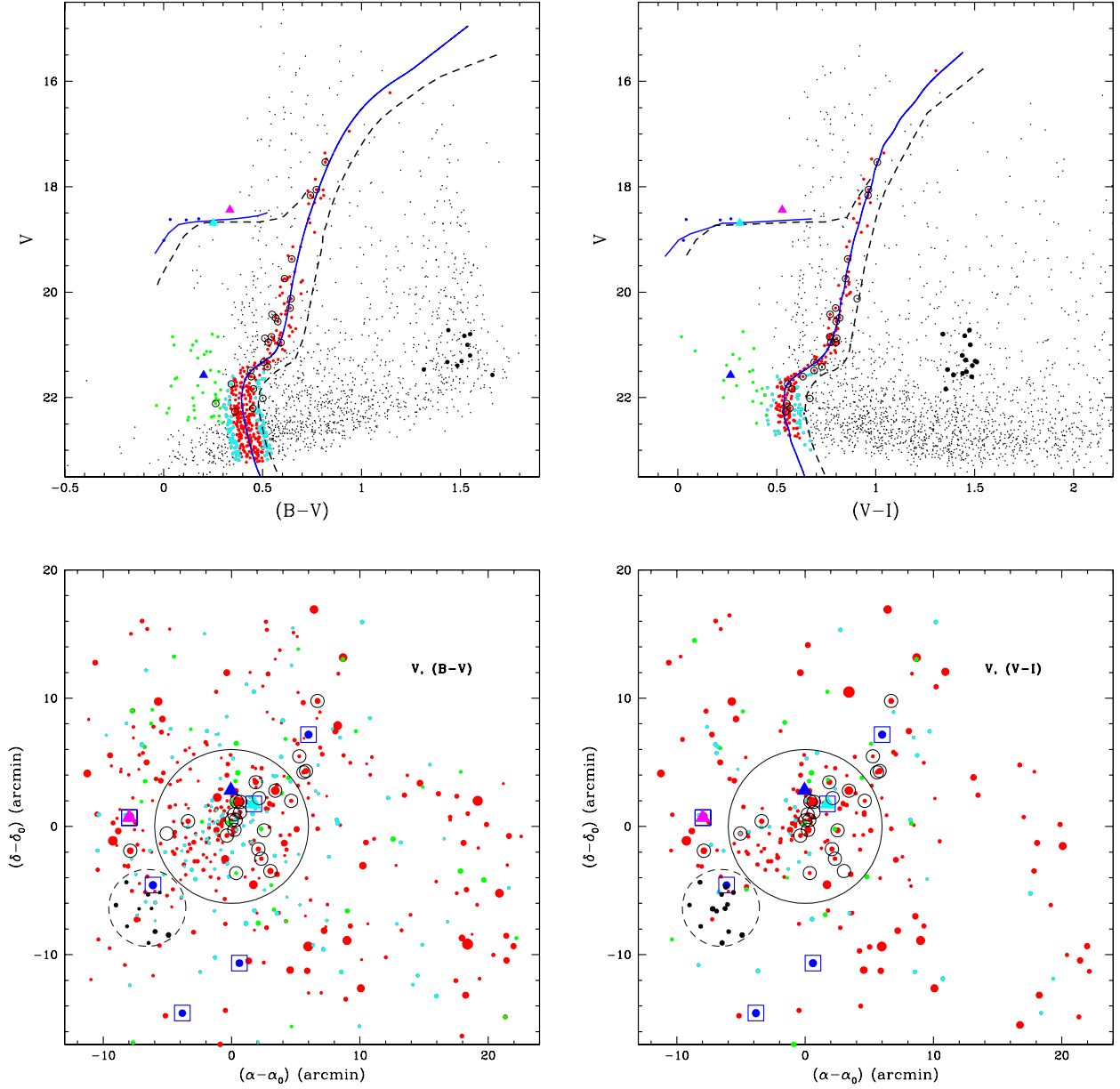}
\caption{{\it Upper panels}: $V$, $B-V$ (left)  and $V$, $V-I$ (right)
CMDs of sources in the INT field of view. Solid (blue) and dashed (black) lines are the ridgelines of the
Galactic GCs M\,68 and M\,3, respectively. Red and cyan dots are stars respectively 
within $\pm$0.06 mag in $B-V$ and  $\pm$0.05 mag in $V-I$, and from $\pm$0.05/0.06 to $\pm$0.1 mag
from the ridgelines of M\,68. 
Blue dots are non-variable stars on the HB, green dots are stars in the BSSs region.  
Variable stars are
marked by filled triangles,  cyan: RRc star, magenta: RRab star, blue: short period variable. Open circles mark 24 member stars of the Coma dSph 
from \citet{ky08}. 
Black dots are galaxies of 
cluster MaxBCG J186.85861+2380004.
{\it Lower panels}: maps of objects most likely belonging to the
Coma dSph, color-coded and with same symbols as in the top-panels of the figure.  HB stars are outlined  by 
squares.
The large circles show the galaxy half-light radius $r_h(exponential)$= 5.9 arcmin (B07).
The dashed circles show the position of the galaxy cluster.
}
\label{f:fig3} 
\end{figure}

The upper panels of Fig.~\ref{f:fig3} show the $V$, $B-V$ and $V$, $V-I$ CMDs of the 
Coma dSph. 
We have plotted all 
stellar-like objects (selected using the $CHI$ and $SHARP$ parameters 
provided by the reduction packages), over the whole 0.5 square degrees
covered by the INT observations. 
This selection is rather secure for
magnitudes above $V \sim$ 20.5-21.0 mag, while it becomes increasingly uncertain
at fainter magnitudes.
The variable stars (filled triangles) are plotted according
to their intensity-averaged magnitudes and colors (see Table~1).
 The CMDs reach $V \sim
23.2$~mag, and appear to be heavily contaminated at every magnitude
level by field objects belonging to the MW halo and disk, and by galaxies which escaped selection. 
We have used the mean ridge lines of the Galactic 
globular cluster M\,68 (NGC\,4590; solid blue lines in Fig.~\ref{f:fig3}), 
drawn from Walker's (1994) $B,V,I$ photometry 
for the cluster, shifted by $\Delta V = +2.95 $~mag in magnitude, and  
$\Delta (B-V)=+0.025$~mag, $\Delta (V-I)=+0.030$~mag in color, to match the horizontal and red giant
branches of the Coma dSph. 
Then we selected as most likely belonging to the galaxy
the sources lying within $\pm 0.06$ mag in $B-V$, and $\pm 0.05$ mag in 
$V-I$, from the ridge lines of M\,68 
(red dots in the upper panels of Fig.~\ref{f:fig3}). The slightly larger range 
adopted for the $B-V$ color takes into account the larger 
uncertainty caused by the small number of $B$ frames. 
To allow for the 
larger photometric errors below $V = 21.5$ mag, we also considered as belonging to the
galaxy stars with colors in the range from $\pm$ 0.05/0.06 to
$\pm$ 0.1 mag from the ridge lines of M\,68  (cyan dots). 
M\,68 is well suited for identifying members of the Coma dSph since,
like Coma, is very metal poor ([Fe/H]$_{\rm M\,68}$= $-2.1$ dex, \citealt{w94}) 
and 
has a well defined and tight red giant
 branch, as well as an extended HB well populated inside and to the blue of the RR Lyrae instability strip (\citealt{w94}). 
 Adopting for M\,68 the reddening value of $E(B-V)$=0.07 $\pm$ 0.01 mag (\citealt{w94})
the color shifts needed to match the horizontal and red giant
branches of Coma imply a
reddening of $E(B-V)=0.045 \pm 0.015$~mag for the galaxy, which can be compared with the value of 
$0.019 \pm 0.026$~mag derived 
from the \citet{sc98} maps. 
 We also show in Fig.~3 results 
 of the fit based on a less metal poor globular cluster: M\,3 
(\citealt{ferraro97}, \citealt{bolte}). 
The red giant branch of M\,3 
is too red and, to match that of Coma, would require for 
the galaxy a negative reddening $E(B-V) \sim -0.06$ mag. With the help of the M\,68 ridgelines 
it is possible to determine  
the average luminosity of
the Coma dSph HB 
roughly 
at the center of the RR Lyrae instability strip: 
$\left<V_{\rm HB}\right>$=18.64 $\pm$ 0.04 mag, and to 
locate the galaxy main sequence turn-off at $V\sim 21.7$
mag. The CMDs also show several stars in the BSS
region (green dots in Fig.~\ref{f:fig3}), which are likely BSSs of the Coma dSph.  
Our identification of the Coma member stars is confirmed in an excellent way by the 
comparison with the \citet{ky08} spectroscopic study of 24 stars with
membership to the Coma dSph confirmed by radial velocity measurements (open 
circles in Fig.~\ref{f:fig3}). The agreement between the two studies  
is 
impressive, and proves the reliability of our procedure to select  
member stars, and our identification of the Coma HB.
The lower panels of Fig.~\ref{f:fig3} show the position of the stars 
we consider to belong to the Coma galaxy in the FOV of the $B$-band (left panel) and $V$-band (right panel) INT observations.
Symbols and color-coding 
are the same as in the top-panels of the figure. Symbol-sizes are proportional to 
the object's magnitudes. The 
HB stars are outlined by squares. 
It is noteworthy that, with the exception only of the RRc variable, all the HB stars 
fall outside the galaxy half-light radius, but generally close to stars whose membership to Coma is confirmed 
by the spectroscopic studies.  

A further feature is visible in the CMDs of the Coma dSph (see upper panels of Fig.~\ref{f:fig3}), 
formed by sources having 
roughly  20.6 $< V <$21.9 mag, 1.3 $ <V-I <1.5$ mag, and $B-V \sim 1.5$~mag. Among them
those marked by black dots are  
mainly concentrated in the left-hand portion of the maps outlined by dashed circles 
(see lower panels of Fig.~\ref{f:fig3}). 
In the literature, at this position, we find  the cluster of galaxies 
MaxBCG J186.85861+2380004 (\citealt{koe07}), which has a photometric
redshift $z=0.243050$. 
The feature we see in the CMDs is, probably, 
the ``red sequence" of this cluster, 
as confirmed by the $V-I$, $B-V$  colors we derive 
for early-type galaxies at z$\sim$0.25 using the GISSEL03 spectral code 
(\citealt{bc03}).

The average magnitude of the Coma dSph HB, set by the matching with the ridge lines 
of M\,68, is 
$\langle V_{\rm HB}\rangle=18.64 \pm 0.04$ mag (where the error includes both uncertainties in the photometry and in the matching procedure).
 Assuming
$M_V$=0.59$\pm$0.03 mag for the absolute magnitude of HB stars at [Fe/H]=$-1.5$ dex (\citealt{cc03}), $\Delta M_V/\Delta
[Fe/H]$=0.214 ($\pm$ 0.047) mag/dex for the slope of the
luminosity-metallicity relation of the RR Lyrae stars (\citealt{cg03}),
$E(B-V)_{\rm Coma}$=0.045 $\pm 0.015$ mag, and [Fe/H]$_{\rm Coma}$=$-2.53 \pm 0.05$ dex (\citealt{ky08}), the
distance modulus of Coma is 18.13 $\pm$ 0.08 mag, ($d$=42$^{+2}_{-1}~\rm{kpc}$).
Errors include uncertainties in the photometry,
reddening, metallicity, and RR Lyrae absolute magnitude. 
On the same assumptions, the average apparent magnitude of the RRab star
$\langle V\rangle =18.44 \pm 0.03$ mag leads, instead, to the distance modulus of  17.93 $\pm$ 0.08 mag, ($d$=39$^{+1}_{-2}~\rm{kpc}$).
While the first estimate is in good agreement with the distance of
$44 \pm 4 $ kpc found by B07, the value inferred from the RRab stars is, respectively, 3 and 5 kpc shorter than the
galaxy distance derived in the present study and in B07.
An additional distance estimate for the two RR Lyrae stars was obtained 
using the empirical Wesenheit relation in the $B, V$ bands as defined by Kov\'acs and 
Walker (2001) and calibrated by  Kov\'acs (2003, this calibration is consistent with 
a distance modulus for the Large Magellanic Cloud of 18.55 mag). The resulting  moduli: $\mu_0 {\rm (RR_{ab})}$ = 18.06$\pm 0.10$ 
and $\mu_0 {\rm (RR_{c})}$ = 18.09$\pm 0.10$ mag, are in good agreement with each other 
and are consistent with the distance inferred from the 
average magnitude of the galaxy HB. The overluminosity of the RRab reflects into the 
Wesenheit index, however it is balanced by the star longer period, thus the 
resulting distance modulus is in good agreement with that derived for the RRc
variable. These results suggest
that the RRab star is brighter 
because it has evolved off the zero age HB, rather than because it is significantly 
closer to us.

\section[]{Summary and conclusions} 

We have identified and obtained $V, I$ light curves for 2 RR~ Lyrae (1
RRc and 1 RRab) and one short period variable,  in the Coma dSph galaxy. The behavior of the RR Lyrae stars suggests
an Oosterhoff II classification for the galaxy.  From the average luminosity
of the HB stars, the distance modulus of the Coma dSph is $\mu_0$=18.13
$\pm$ 0.08 mag ($d=42^{+2}_{-1}~\rm{kpc}$). 
%
The galaxy appears
to be elongated with irregular and extended shape, likely caused by the tidal interaction with the
MW. 
The RRab star lies outside the galaxy half-light radius and is about 0.2 mag brighter than the average magnitude
of the galaxy's HB. Its bright mean magnitude leads to a distance of $d$=39 $^{+1}_{-2}~\rm{kpc}$, which, 
if interpreted in terms of projection effects, could  provide a rough estimate for the galaxy elongation. However, 
the Wesenheit results suggest that the star overluminosity is indeed due, at least 
in part, to 
evolution off the zero-age HB.
Finally, from the analysis of the CMDs we identify in the
lower north-east field of the Coma galaxy, the background cluster of galaxies MaxBCG J186.85861+2380004
(\citealt{koe07}).

\acknowledgments
We warmly 
thank Dr Evan Kirby %
for sending us the identification and 
individual metallicities for member stars of the
Coma dSph galaxy, Dr Amata Mercurio for advice with the identification of the galaxy 
cluster, and the anonymous referee for helpful comments. 
%
Financial support for this study was provided 
by Regione Campania 2003 (PI J. Alcal\`a) and by PRIN-INAF 2006 (PI
G. Clementini). HAS thanks the US NSF for support under grant AST 0607249.

\clearpage

  \begin{table*}
  \tiny
      \caption[]{Identification and properties of variable stars in the Coma dSph
      galaxy}
         \label{t:table1}
     $$
         \begin{array}{lccllcccrrc}
	    \hline
            \hline
           \noalign{\smallskip}
           {\rm Name} &  {\rm \alpha } & {\rm \delta} &  {\rm Type} &~~~~P & 
	    ~~~{\rm Epoch (max)}  & \langle V\rangle  & \langle I\rangle  & A_V~~ 
	    & A_I~~&{\rm [Fe/H]_{ZW}} \\
            ~~ & {\rm (2000)}& {\rm (2000)}& & ~{\rm (days)}& ($-$2450000) & {\rm (mag)} & {\rm (mag)} 
	     & {\rm (mag)} &  {\rm (mag)}& \tablenotemark{a} \\
            \noalign{\smallskip}
            \hline
            \noalign{\smallskip}
	    
{\rm ~V1} &12~27~33.50&23~54~55.7& {\rm RRab }      &0.66971 &4171.583 &18.44 &17.74&0.78&0.53&$-2.068$\\ 
{\rm ~V2} &12~26~50.89&23~56~00.6& {\rm RRc }       &0.31964 &4172.128 &18.69 &18.20&0.37&0.24&$-2.074$\\ 
{\rm ~V3} &12~26~58.68&23~57~04.5& {\rm Short~period}    &0.12468 &4171.950 &21.57 &21.13&0.29&0.28&$\nodata$\\ 
\hline
            \end{array}
	    $$
\tablenotetext{a}{Metallicities derived from the Fourier parameters
of the light curves.}
\end{table*}

\end{document}